\documentstyle[11pt,lanl,epsfig]{article}

\def\prl{\em Phys. Rev. Lett.}

\def\np#1#2#3 {Nucl. Phys. {\bf#1},{\ #2} (19#3)}
\def\prev#1#2#3 {Phys. Rev. {\bf#1},{\ #2} (19#3)}
\def\be{\begin{equation}}
\def\ee{\end{equation}}
\def\bea{\begin{eqnarray}}
\def\eea{\end{eqnarray}}

\def\gw{\mbox{$\Gamma_W$}}
\def\pbarp{\mbox{$\overline{p}p$}}
\def\cme{\mbox{$\sqrt{s}=1.8\;\rm TeV \ $}}
\def\Wtn{\mbox{$W \rightarrow \tau \nu_\tau \ $}}
\def\Ztt{\mbox{$Z \rightarrow \tau \tau \ $}}
\def\Zll{\mbox{$Z \rightarrow l l \ $}}
\def\Wen{\mbox{$W \rightarrow e \nu \ $}}
\def\Zee{\mbox{$Z \rightarrow e e \ $}}
\def\Zeemiss{\mbox{$Z \hskip -0.1cm \rightarrow \hskip -0.1cm ee$}}
\def\Wmn{\mbox{$W \rightarrow \mu \nu_\mu \ $}}
\def\Zmm{\mbox{$Z \rightarrow \mu \mu \ $}}
\def\ipb{pb$^{-1}$}
\def\Wln{\mbox{$W \rightarrow l \nu_l \ $}}
\def\stw{\mbox{$\sigma(\pbarp \to W + X)$}}
\def\stz{\mbox{$\sigma(\pbarp \to Z + X)$}}
\def\sw{\mbox{$\sigma_W$}}
\def\sz{\mbox{$\sigma_Z$}}
\def\glw{\mbox{$\Gamma(W \to l \nu)$}}
\def\glz{\mbox{$\Gamma(Z \to l^+ l^-)$}}
\def\gz{\mbox{$\Gamma_Z$}}
\def\D0{D$\!$\O }
\def\etal{{\it et al.}}
\begin{document}
\lefthyphenmin=2
\righthyphenmin=3
\vspace*{4cm}
\title{\boldmath
REVIEW OF $W$ AND $Z$ PRODUCTION AT THE TEVATRON}

\author{ Cecilia E.\ GERBER }

\address{Fermi National Accelerator Laboratory\\
        P.O. Box 500, Batavia, IL 60510, USA\\
	%\rm gerber$@$fnal.gov\\
	\rm (for the \D0 and CDF Collaborations)}

\maketitle\abstracts{The CDF and \D0 collaborations have used recent data taken
at the Tevatron
 to perform QCD tests with $W$ and $Z$ bosons decaying leptonically.
\D0 measures the production cross section times branching ratio for $W$ and $Z$
bosons. This also gives an
indirect measurement of the total width of the $W$ boson:
$\gw=2.126\pm0.092\;\rm GeV$. CDF reports on a direct measurement of
$\gw=2.19\pm0.19\;\rm GeV$, in good agreement with the indirect determination
and Standard Model predictions.
\D0's measurement of the differential $d\sigma/dp_T$ distribution for $W$ and
$Z$ bosons decaying to electrons agrees with the combined QCD perturbative and
resummation calculations. 
In addition, the $d\sigma/dp_T$ distribution for the $Z$ boson discriminates
between different vector boson production models.
Studies of $W+\;\rm Jet$ production at CDF find the NLO QCD prediction for 
the production rate of $W+\ge1\;\rm Jet$ events to be in good agreement with the
data.} 

\section{Introduction}

$W$ and $Z$ bosons, the carriers of the weak force, are directly produced in
high energy $\pbarp$ collisions at the Fermilab Tevatron, which operates at a
center of mass energy of $\cme$. 
In addition to probing electroweak physics, the study of the
production of $W$ and $Z$ bosons 
provides an avenue to explore QCD, the theory of strong
interactions.  The benefits of using intermediate vector bosons to
study perturbative QCD are  large momentum transfer, distinctive
event signatures, low backgrounds, and a well understood electroweak
vertex.

Large numbers of $W$ bosons have been
detected by the two collider detectors (CDF and \D0), during the 1992--1996
running period. These samples complement the detailed
studies carried out on the $Z$ boson at LEP and SLC, and also the new $W$ 
studies from LEP II.
The CDF and \D0 collaborations have used these data 
to perform various tests of the Standard Model. The
preliminary results are presented in the next sections.

\section{\boldmath
$W$ and $Z$ Production Cross Sections at \D0 }
\label{sec:wzxsec}

The $W$  and $Z$  production  cross  sections times
leptonic branching  fractions are measured using data collected by 
the \D0 detector during 1994--1995, in the electron, muon and tau channels.
The $W$ candidates decaying to electrons or muons were selected as 
events containing one high quality isolated lepton and an imbalance of the
momentum in the transverse plane of at least $25\;\rm GeV$ for electrons and
$20\;\rm GeV$ for muons, as a signal for the undetected neutrino. The $Z$
candidates decaying to electrons or muons were selected as
events containing two high quality isolated leptons. 

The major source of 
background in all four cases is due to QCD events with jets faking leptons.
The amount of background in the samples is
estimated directly from collider data.
Backgrounds originating from other physics processes ($\Wtn,\Ztt$, Drell--Yan) 
are estimated from Monte Carlo. 
Lepton selection efficiencies are determined
from the $\Zll$ data. The geometric and kinematic acceptance is calculated
from a fast Monte Carlo simulation of the \D0 detector.
Table~\ref{tab:xsec} shows the preliminary values for the cross sections 
measured from these samples.

\begin{table*}[hbt]
\setlength{\tabcolsep}{1.5pc}
\newlength{\digitwidth} \settowidth{\digitwidth}{\rm 0}
\catcode`?=\active \def?{\kern\digitwidth}
\caption{The \D0 preliminary cross sections for $W$ and $Z$ bosons.}
\label{tab:xsec}
\begin{tabular*}{\textwidth}{@{}l@{\extracolsep{\fill}}cccc}
\hline
                 & \multicolumn{1}{r}{$\Wen$}
                 & \multicolumn{1}{r}{$\Zee$}
                 & \multicolumn{1}{r}{$\Wmn$}
                 & \multicolumn{1}{r}{$\Zmm$}         \\
\hline
$N_{obs}$        &  59579     & 5705      & 10335      & 331  \\
Background($\%$)  & $8.1\pm0.9$ &$4.8\pm0.5$ & $19.8\pm1.9$ & $11.6\pm2.3$\\
Efficiency($\%$)  & $70.0\pm1.2$ & $75.9\pm1.2$ & $26.0\pm1.6$ & $51.9\pm3.6$\\
Acceptance($\%$)  & $43.4\pm1.5$ & $34.2\pm0.5$ & $20.5\pm0.2$ & $4.9\pm0.2$\\
Luminosity [\ipb] & $75.9\pm6.4$ & $89.1\pm7.5$ & $65.3\pm3.5$ & $65.3\pm3.5$\\
\hline
$\sigma \cdot B$ [nb] ($\pm$stat) & $2.38\pm0.01$ & $0.235\pm0.003$ &
                                    $2.38\pm0.03$ & $0.176\pm0.011$ \\
($\pm$syst) ($\pm$lum)            & $\pm0.09\pm0.20$ & $\pm0.005\pm0.020$ &
                                    $\pm0.17\pm0.13$ & $\pm0.020\pm0.009$ \\
\hline
\end{tabular*}
\end{table*}

Many common sources of systematic error cancel when taking the ratio of the
$W$ to $Z$ production cross section times branching ratio, defined as
\[R \equiv \frac{\stw B(\Wln)}{\stz B(\Zll).} \]
This ratio is of interest since it can be expressed as the product of 
calculable or well measured quantities:
\[R= \frac{\sw}{\sz} \hskip3mm \frac{\glw}{\glz} \hskip3mm \frac{\gz}{\gw}.\]
For the combined electron and muon channels, \D0 measures
\[R = 10.48  \pm 0.43.\]
Using the LEP measurement~\cite{PDG} of $B(\Zll) =  (3.367\pm0.006)\%$,
and the theoretical
calculation~\cite{theory} of $\sigma_W/\sigma_Z  = 3.33 \pm  0.03$, one obtains
\[B(\Wln) = (10.59 \pm 0.44) \%. \]
Combining this result with a theoretical calculation of the $W$ leptonic partial
width~\cite{width} $\Gamma (\Wln) = 225.2 \pm 1.5$~MeV, results in a 
total width for the $W$ boson of
\[\Gamma (W) = 2.126 \pm 0.092 \mbox{\ GeV}.\]
This method, though model dependent,  gives the most
precise indirect measurement of the width of the $W$ boson ($\gw$) currently
available.
%, with the uncertainty expected to reduce when the final systematic
%studies are completed.
All these results are in good
agreement with previous \D0\ results~\cite{d01axsec}, and with Standard Model
predictions~\cite{width}.

\D0 has also observed the production of $\Wtn$ and used it to test lepton
universality. The $\tau$ lepton is identified via its 
hadronic decay, which is detected as an 
isolated, narrow jet, with $E_T(jet)>25\;\rm GeV$ and completely contained 
within \D0's central calorimeter. In addition, to select $\tau's$ originating 
from $W$ decays, a minimum imbalance in the transverse energy of $25\;\rm GeV$ 
is required. 
The Profile, defined as the sum of the two highest $E_T$ towers divided by the
transverse energy of the cluster, provides powerful discrimination against QCD
multijet 
backgrounds. Jets originating from hadronic $\tau$ decays tend to be
narrower than those originating in multijet QCD events, and therefore
will show a higher value for the Profile distribution. The
low--Profile region is used to estimated the QCD contamination in the final
$\Wtn$ sample.

Events where more than one inelastic collision took place during the same
beam crossing were rejected at the trigger level; this
effectively reduced the integrated luminosity to $\approx 17\;\rm pb^{-1}$ for
the complete 1994--1995 \D0 data sample.
1202 events pass these selection criteria, with estimated backgrounds of
$106\pm7\pm5$ events from QCD, $81\pm14$ events from noisy calorimeter cells,
$32\pm5$ events from $\Ztt$, and $3\pm1$ events from $\Wen$. 
The acceptance $\times$ efficiency for the selection is $\approx 3.8\%$. 
The preliminary cross section times branching ratio obtained from this data
is 
\[\stw B(\Wtn) = 2.38\]
\[\pm0.09\;{\rm (stat)} \pm0.10\;{\rm(syst)}\pm0.20\;\rm(lum) nb.\]
 Comparing this measurement with \D0's
published~\cite{d01axsec} value for $\sigma \cdot B(\Wen)$, measures the
ratio of the couplings 
\[g_{\tau}^W/g_e^W=1.004\pm0.019\pm0.026.\]
 This result
shows good agreement with the expected $e-\tau$ universality.

\section{\boldmath
Direct Measurement of the Width of the $W$ at CDF }
The indirect measurement of $\gw$ presented in the previous section assumes that
the $W$ coupling to leptons is given by the Standard Model. 
Although in principle
nonstandard couplings would also alter the $W$ production cross section and
thus affect the value of $\Gamma(W\to l\nu)/\gw$ extracted from $R$, a direct
measurement of the total width of the $W$ is desirable so that these radiative
corrections to $\gw$ can be observed.

It has been shown~\cite{cdfgw1A}  that the tail of the transverse mass $M_T$
distribution of the $W$ contains information on $\gw$. Events with $M_T > M_W$
can arise due to the nonzero $W$ width or due to the calorimeter resolution.
However, a precise measurement of $\gw$ from the high mass tail is possible
because far above $M_W$ the Breit--Wigner tail dominates over the Gaussian
resolution of the detector. In this analysis CDF uses data taken during the
1994--1996 Tevatron run to determine the $W$ width from a binned log--likelihood
fit to the transverse mass distribution in the region 
$110<M_T(\;{\rm GeV}/c^2)<200$. 

\begin{figure}[t]
\centerline{\hbox{
\psfig{figure=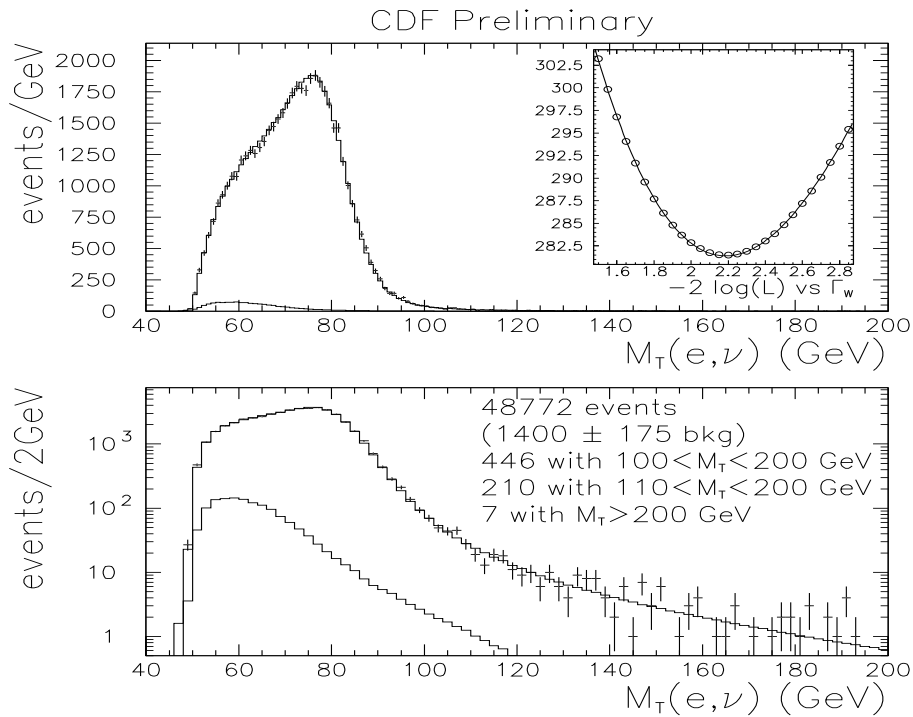,width=12cm,height=9.5cm}}}
\caption{CDF $W$ data transverse mass, with best fit overlayed. The size and
shape of the summed backgrounds are also shown. Inset: Log--likelihood fit of
the data to Monte Carlo templates of different $W$ widths.}
\label{fig:cdfgw}
\end{figure}

Monte Carlo templates are generated for different
values of $\gw$ and correspond to the sum of the $W$ Monte Carlo and the
backgrounds. The data are fitted to each template, and a likelihood curve vs
$\gw$ is made. The resulting likelihood is shown in the inset to
figure~\ref{fig:cdfgw}. The most likely value for the total width of the $W$
boson is $\gw=2.19^{+0.17}_{-0.16}\rm(stat)\pm0.09(syst)\;\rm GeV$.
This result is in good agreement with the indirect measurement and the Standard
Model prediction~\cite{width}.
\newpage

\section{\boldmath
Measurement of the differential $d\sigma/dp_T$ Cross Sections at \D0}

\D0 also measures the differential cross section for the $W$ and the $Z$ boson
decaying to electrons, as a function of the boson transverse momentum. 
The transverse momentum ($p_T$) of intermediate vector bosons
    produced in proton-antiproton collisions is due to the production of
    one or more gluons or quarks along with the boson.
    At low transverse momentum ($p_T < 10\;\rm GeV/\it c$),
    multiple soft gluon emission is expected to dominate the
    cross section. A soft gluon resummation 
    technique~\cite{CSStheory,DWStheory,AEGMtheory,ly,ak} is therefore 
    used to make QCD predictions.  At high transverse momentum
    ($p_T  > 20\;\rm GeV/\it c$), the cross section is dominated by the
    radiation of a single parton with large transverse momentum.
    Perturbative QCD~\cite{ar} is therefore expected to be reliable in
    this regime.  A prescription~\cite{ak}
    has been proposed for matching
    the low and high $p_T$ regions to provide a
    continuous prediction for all $p_T$.
    Thus, a measurement of the transverse momentum distribution may be used to
    check the soft gluon resummation calculations in the low $p_T$
    range, and to test the perturbative QCD calculations at high $p_T$.

\begin{figure}[htb]
\centerline{\hbox{{\epsfysize=8cm \epsffile{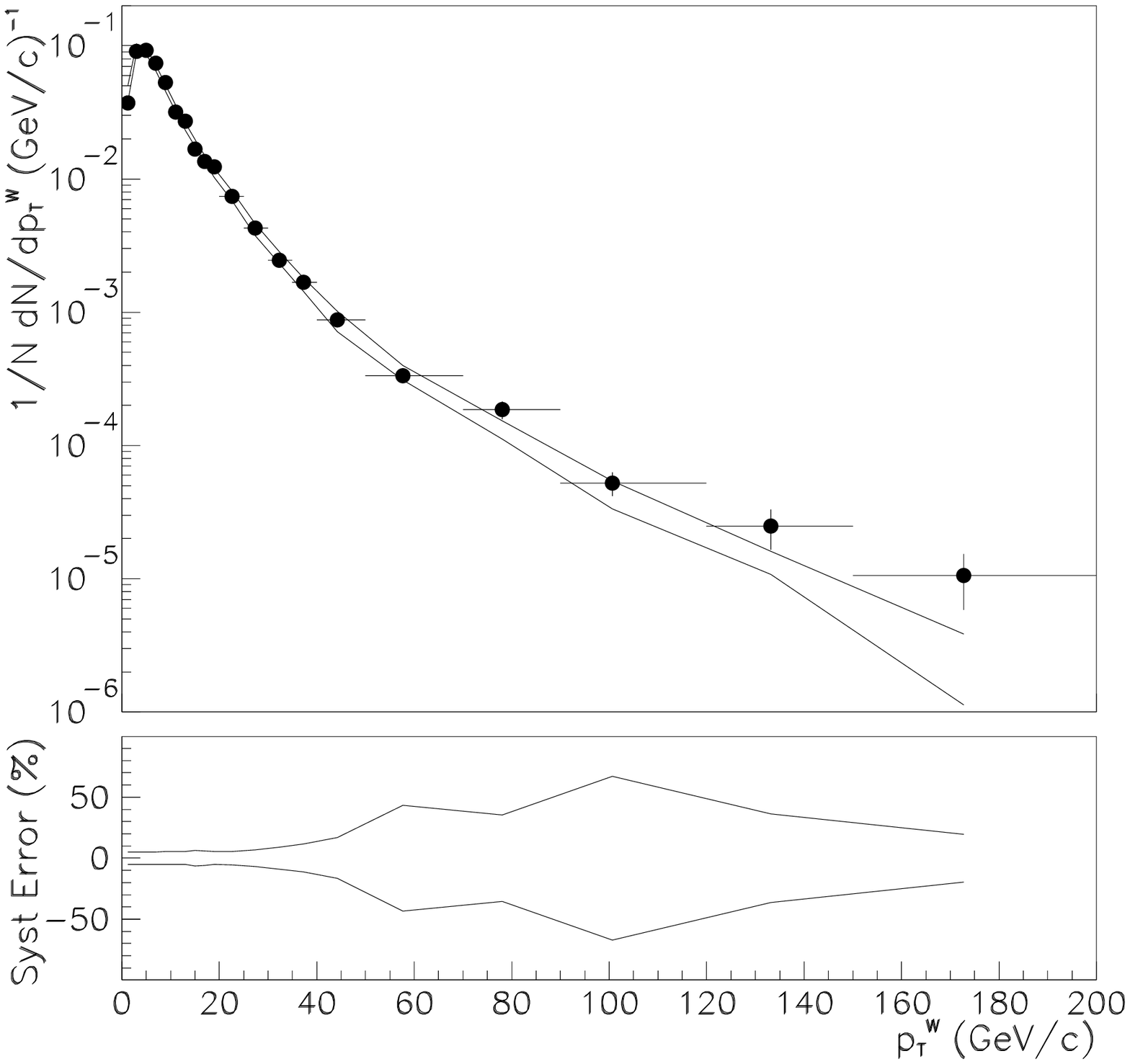}}
{\epsfysize=8cm \epsffile{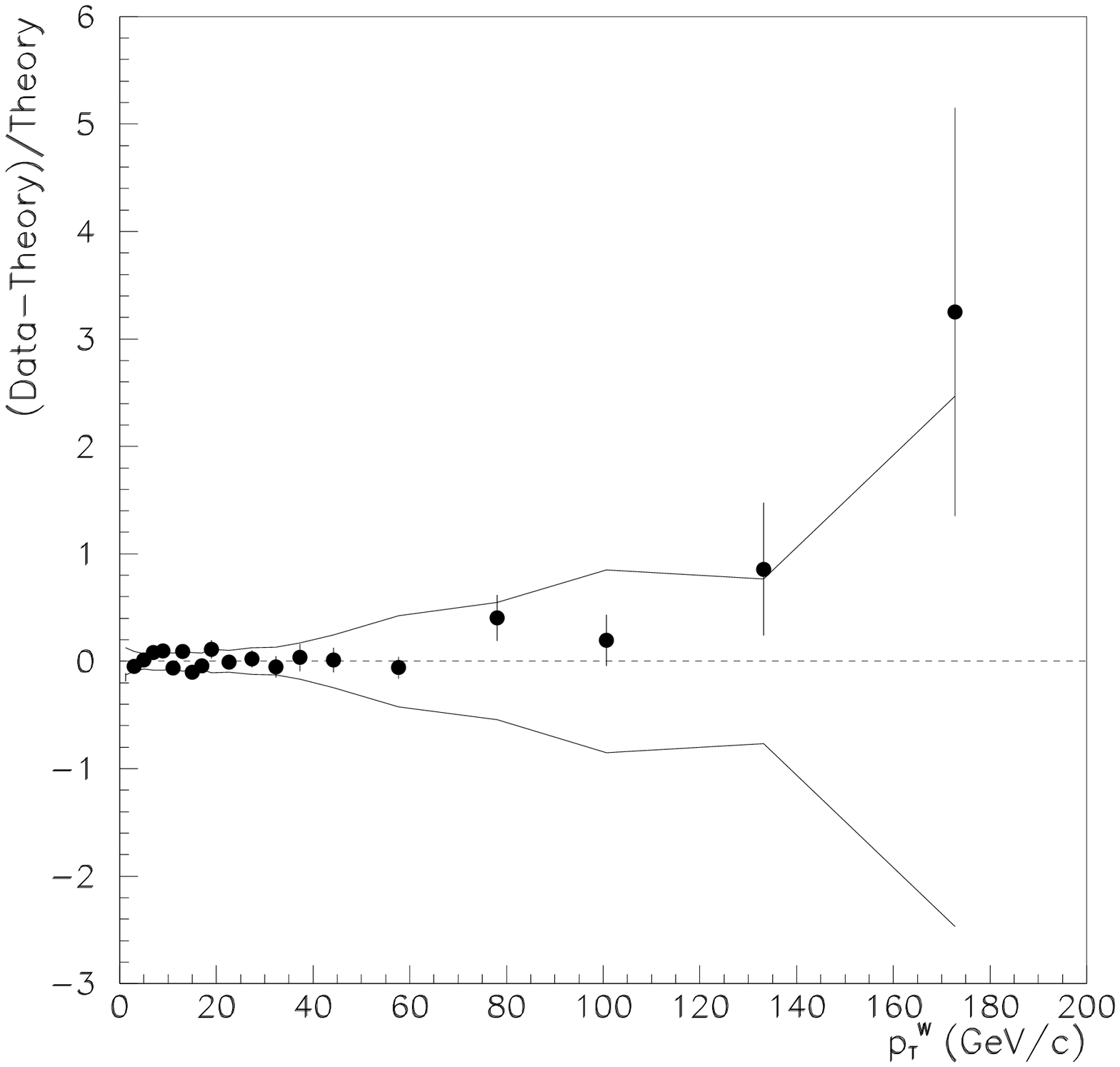}}}}
\caption{Left: \D0 data (solid points) with statistical uncertainty are compared
to the theoretical prediction by Arnold--Kauffman[10], smeared for detector
resolutions. Data and theory are independently area--normalized to unity. 
The fractional systematic uncertainty on the data
is shown as a band in the lower portion of the plot.
Right: The ratio (Data--Theory)/Theory shown as a function of
$p_T^{W}$ with its statistical uncertainty as error bars.}
\label{fig:wpt}
\end{figure}

\begin{figure}[htb]
\centerline{\hbox{{\epsfysize=8cm \epsffile{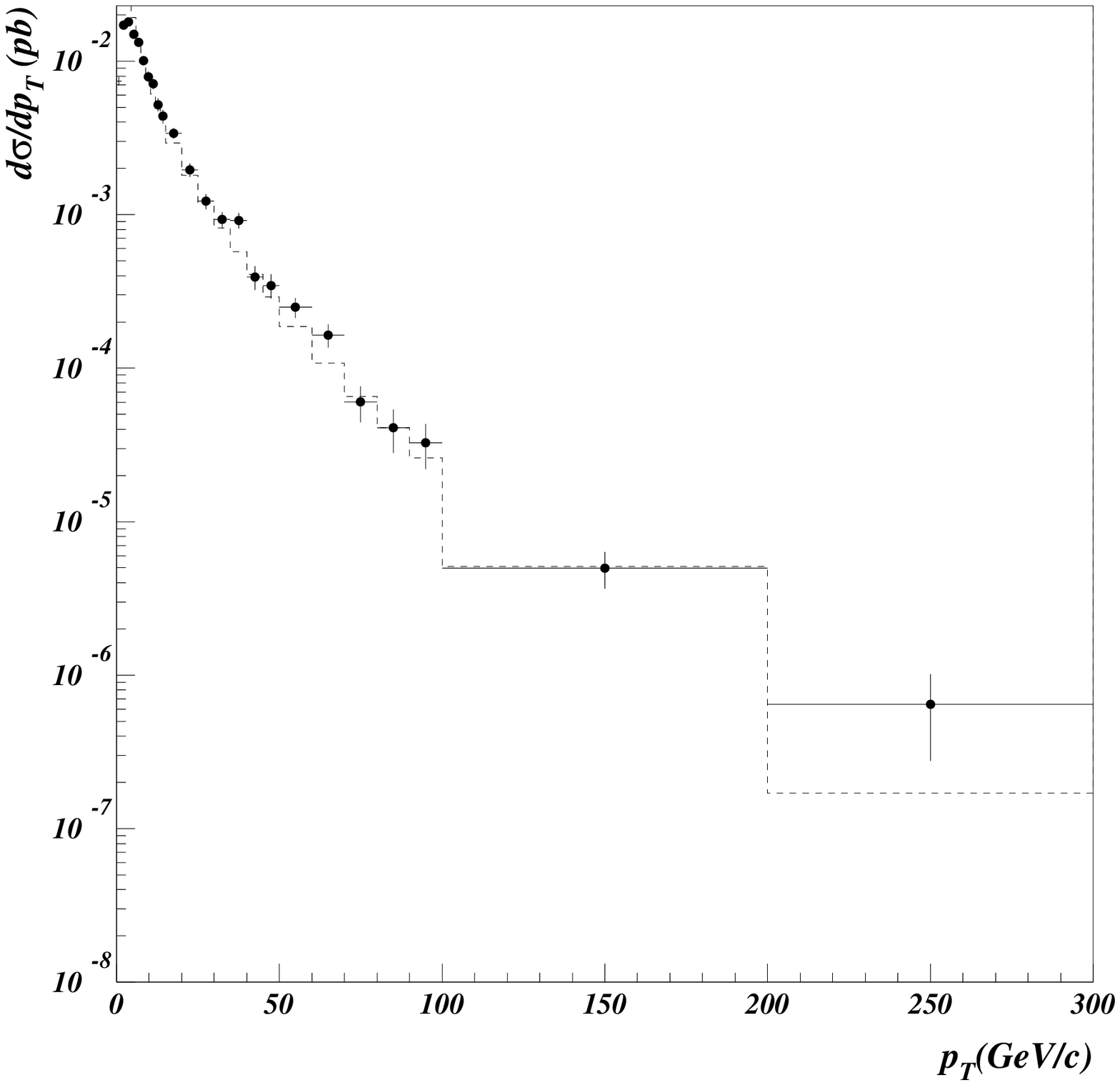}}
{\epsfysize=8cm \epsffile{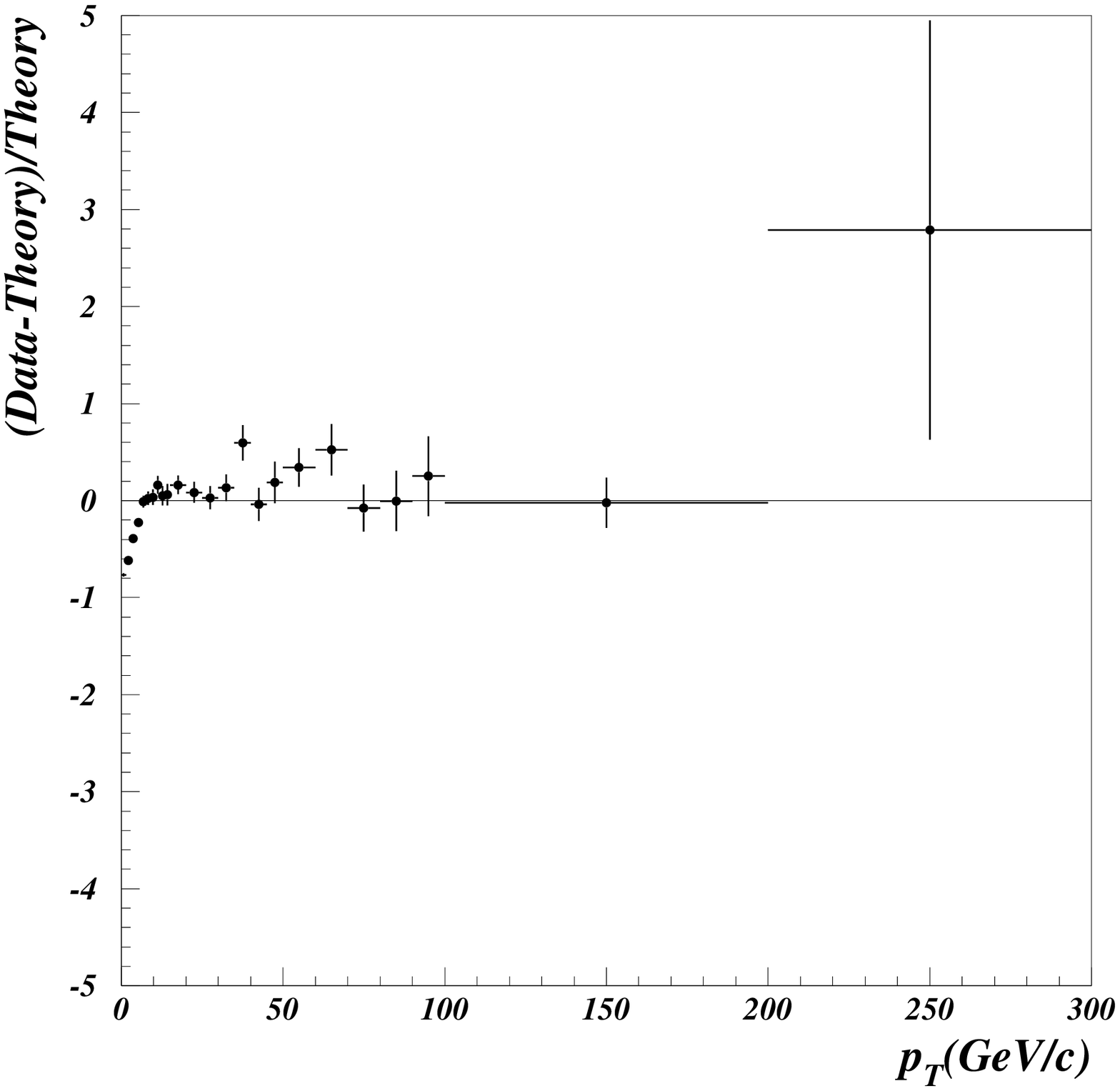}}}}
\caption{Left: \D0 data (solid points) with total uncertainty are compared
to the theoretical prediction by Arnold--Reno[11], smeared for detector
resolutions. Data is normalized to the measured inclusive $Z$ production cross
section presented in section~\ref{sec:wzxsec}; the theory is normalized to its
own prediction. Right: The ratio (Data--Theory)/Theory as a function of
$p_T^{Z}$ clearly shows the known need for resummation in the very low $p_T$
region.} 
\label{fig:zpt}
\end{figure}

For this analysis, \D0 uses 7132 $\Wen$ events collected during the 1992--1993
collider run, and 6407 $\Zee$ events from the 1994--1996 data set. The major
source of background for both samples is QCD multi--jet and photon--jet events:
the amount of background in the samples and its shape as a function of $p_T$ is
obtained directly from \D0 data. The variation of the
trigger and selection efficiency vs $p_T$ is calculated using a full detector
Monte Carlo simulation. A parametrized representation of the D\O\ detector
is used to smear the theoretical prediction by detector effects and
compare it to the measured $p_T$. The results are shown in figures~\ref{fig:wpt}
and \ref{fig:zpt} for the $W$ and $Z$ respectively. The $W$ data shows good
agreement with the combined QCD perturbative and resummation calculation.
The $Z$ data shows generally good agreement over the entire range, except for an
approximately $2\sigma$ deviation near $p_T^Z$ of $40\;\rm GeV/{\it c}$.
In addition, figure~\ref{fig:zpt} shows
how the $p_T^Z$ distribution departs from
the pure perturbative QCD prediction at very low transverse momenta, which
corroborates the known need for resummation in this region.

\begin{figure}[htb]
\centerline{\hbox{
\psfig{figure=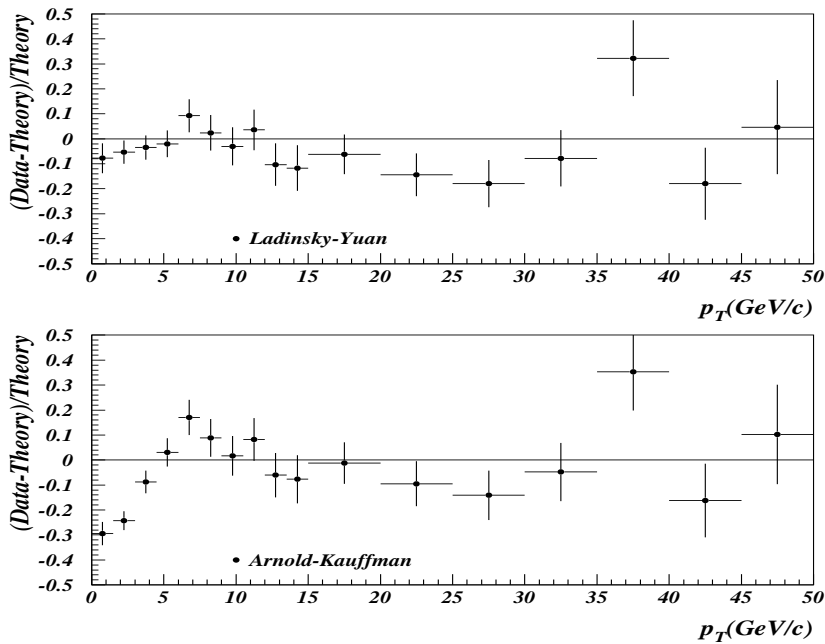,width=12cm,height=9.5cm}}}
\caption{ The ratio (Data--Theory)/Theory shown as a function of
$p_T^{Z}$ for two available models for the non--perturbative contribution:
Ladinsky--Yuan [9] (top), and Arnold--Kauffman [10] (bottom).}
\label{fig:lowptz}
\end{figure}

\clearpage
Figure~\ref{fig:lowptz} shows that the $p_T^Z$ distribution is able to
distinguish between two available models~\cite{ly,ak} for the non--perturbative
contributions to the transverse momentum.

\section{\boldmath
Studies of  $W+\;\rm Jets$ Production at CDF}

The CDF Collaboration uses $108\;\rm pb^{-1}$ of data, accumulated from 1992 to
1995, to measure the production cross section of 
$\sigma_n \equiv W+\ge n\;\rm Jets$ (for $n=1,4$) relative to the published
($W+\ge 0\;\rm Jets$) inclusive cross section measurement  
for the 1992--1993 collider run~\cite{xsec_cdf}. The analysis is
restricted to events where the $W$ decays into a central pseudorapidity 
electron ($|\eta_e|<1.1$). The jets are reconstructed using a 0.4 fixed cone
algorithm and restricted to $|\eta_{\rm Jet}|<2.4$. 

The backgrounds to this sample
are determined as a function of jet multiplicity in the event. The dominant
background is due to QCD multijet events, and varies from $\approx 2.9\%$ to
$\approx 27\%$ for the $n=0$ to $n=4$ case. The background originating from 
top quark and diboson production processes varies from $\approx 0.1\%$ to
$\approx 17\%$ for the $n=0$ to $n=4$ case. Backgrounds from
electroweak processes ($\Wtn$, $\Ztt$ and $\Zeemiss$) are estimated as a flat
$3\%$ contribution. The overall detection efficiency is $\approx 20\%$.

The measurement~\cite{cdfwjets} 
of $\sigma_n$ is compared to predictions obtained from
the leading order QCD matrix element calculations~\cite{vecbos} 
by including gluon radiation 
and hadronic fragmentation using the HERWIG~\cite{herwig} shower simulation
algorithm. The $W$ boson events with hadron showers are then introduced into
a full CDF detector simulation, and the resulting jets are identified and 
selected as in the data, allowing a comparison between the QCD 
predictions and the data. 

\begin{figure}[htb]
\centerline{\hbox{
\psfig{figure=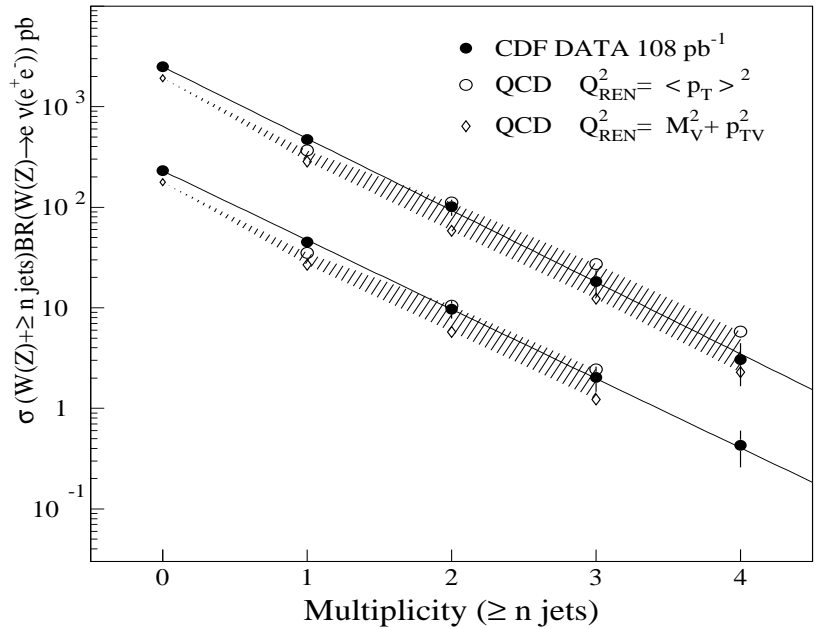,width=12cm,height=9.5cm}}}
\caption{CDF's $\sigma_n(W+n\;\rm Jets)$ as a function of the jet
multiplicity.  
The result[12] for $\sigma_n(Z+n\;\rm Jets)$ is also shown.}
\label{fig:cdf_xsec}
\end{figure}

Figure~\ref{fig:cdf_xsec} shows the CDF data compared to the theoretical
prediction for two choices of the renormalization and factorization scale
$Q^2_{REN,FAC}$.
The published $Z+\ge n\;\rm Jets$ cross section~\cite{cdfzjets} 
is also shown. One observes that the measured $W+\ge n\;\rm Jets$ cross
sections are a factor of 1.7 larger than the LO QCD prediction for 
$Q^2_{REN,FAC}=M^2(W)+p_T^2(W)$.

CDF uses the same data sample to measure
$R_{10} \equiv \sigma(W+\ge1\;{\rm Jet})/\sigma(W)$,
as a function of jet $E_T$ threshold ($E_T^{min}$) in the 15 to 95 GeV range.
The backgrounds to the $W+\ge1\;\rm Jet$ sample
are determined as a function of $E_T^{min}$.
The dominant
background is due to QCD multijet events and varies from 
$\approx 13\%$ to $\approx 28\%$ for $E_T^{min}$ of 15 to 95 GeV.
The corresponding backgrounds originating from other physics
processes vary from $\approx 5\%$ to $\approx 13\%$.
The combined selection efficiency and acceptance varies from 
$\approx 19\%$ to $\approx 26\%$ as a function of jet $E_T^{min}$.
For the inclusive $W$ sample the total background level is $\approx 6\%$, and
the combined selection efficiency and acceptance is $\approx 20\%$.

\begin{figure}[htb]
\centerline{\hbox{{\epsfysize=8cm \epsffile{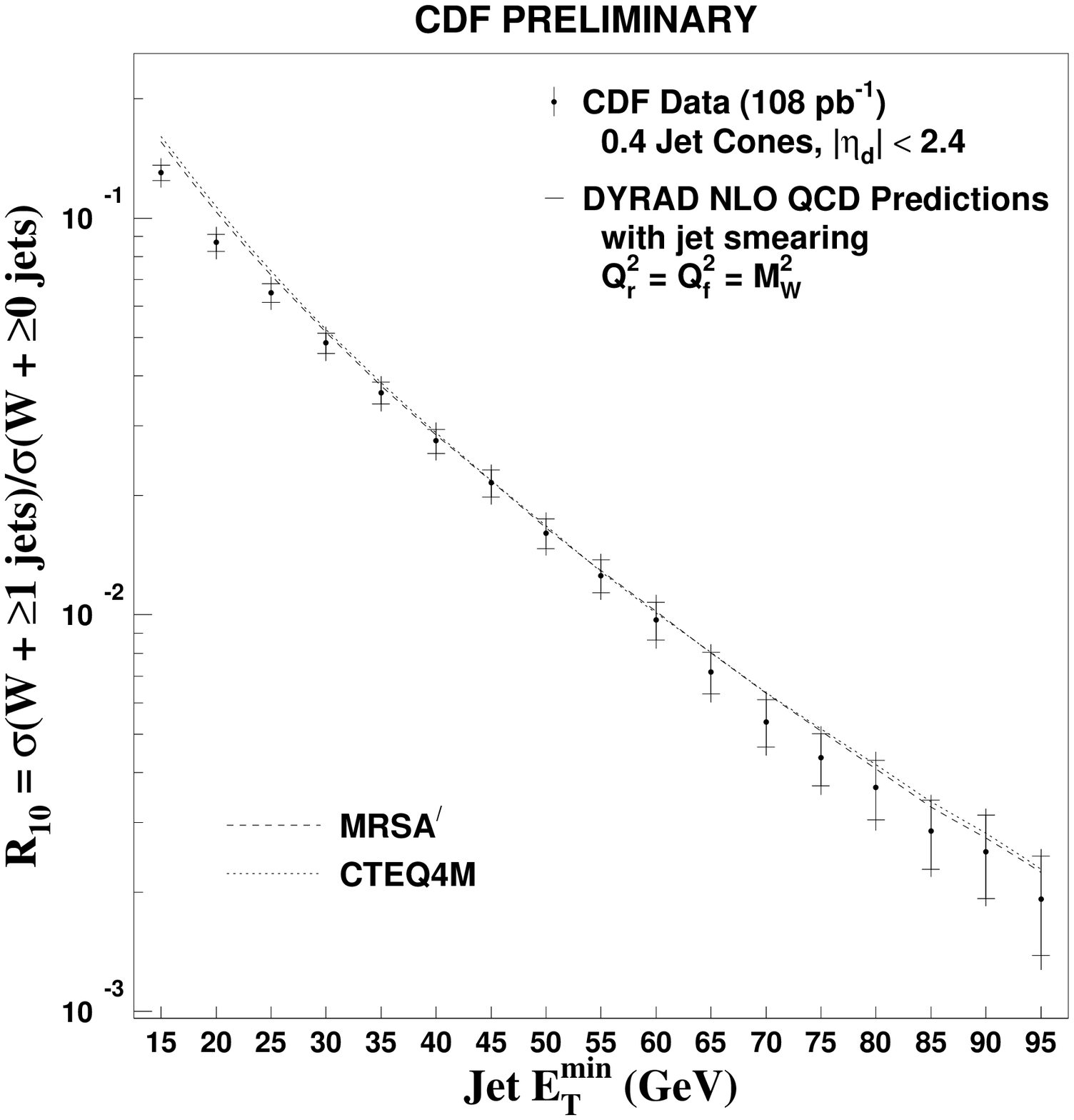}}
{\epsfysize=8cm \epsffile{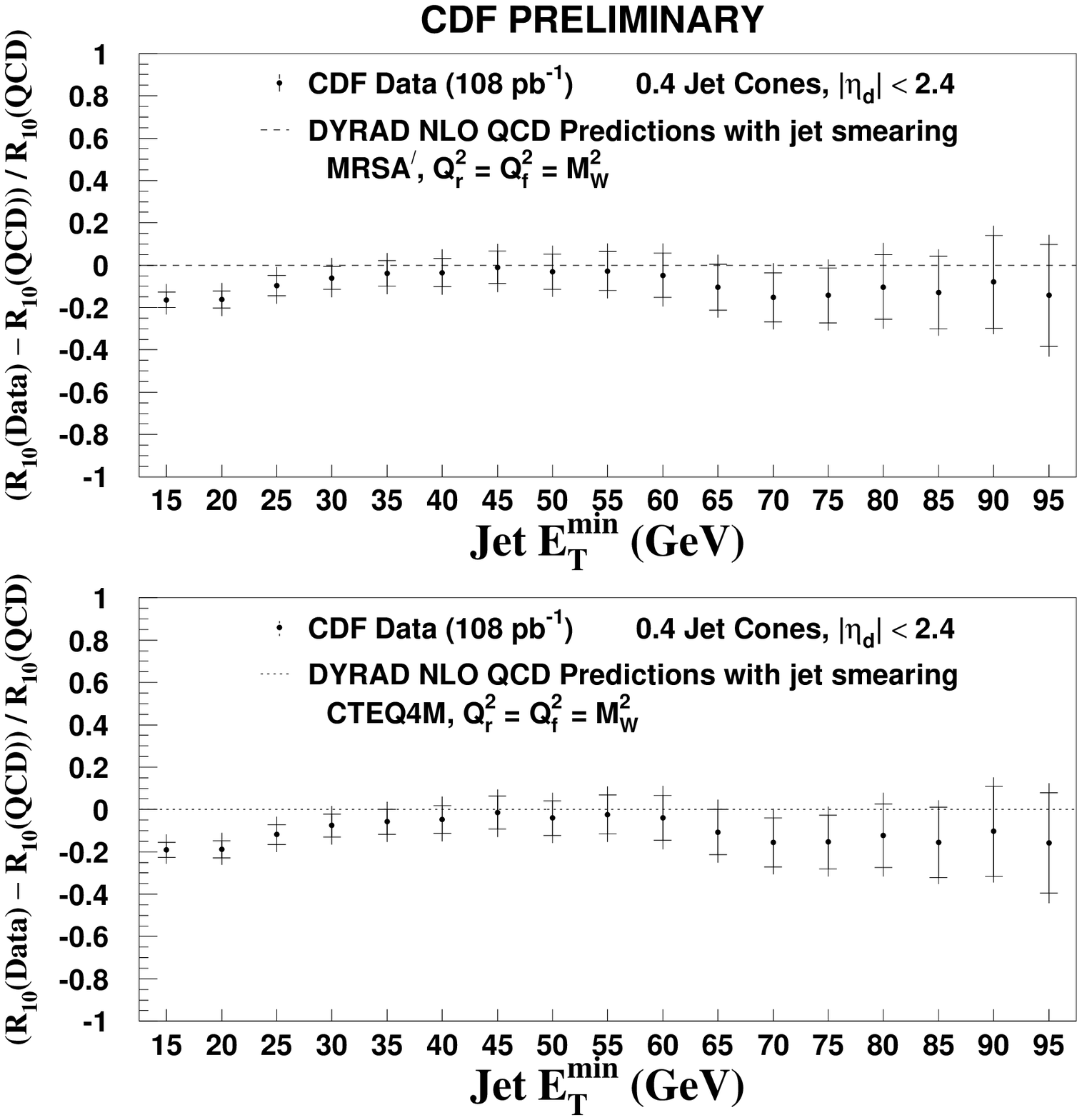}}}}
\caption{CDF's measurement of 
$R_{10} \equiv \sigma(W+\ge1\;{\rm Jet}/\sigma(W)$,
as a function of jet $E_T^{min}$, 
compared to NLO QCD prediction by[17]. The (Data--Theory)/Theory plot is shown
at the right for two choices of PDFs.}
\label{fig:cdfr10}
\end{figure}

The measurement of $R_{10}$ is compared to NLO QCD predictions obtained by the
DYRAD Monte Carlo~\cite{dyrad}.
Figure~\ref{fig:cdfr10} shows the measured $R_{10}$ distribution as a
function of $E_T^{min}$, compared to the theoretical prediction.
The agreement over the full Jet $E_T^{min}$ range is excellent, except for the
lowest thresholds ($E_T^{min}<25\;\rm Gev$), which is interpreted as an
indication of the need of resummation in the calculation of the cross sections.
Little variation is observed in the prediction when varying parton
distribution functions (PDFs) or the normalization and factorization scales.

\section{Conclusions}

Using approximately $100\;\rm pb^{-1}$ of data each from Run 1 at the Tevatron,
the CDF and \D0 collaborations make precise measurements of the vector boson
production properties that test QCD predictions. 
The measurements of the inclusive $W$ and $Z$ production cross sections agree
with Standard Model predictions. The direct and indirect determinations of the
total width of the $W$ boson agree with each other and with expectations.
The measured $W$ and $Z$ transverse momentum distributions agree with the
combined QCD perturbative and resummation calculations; the $Z$ boson data
distinguish between different vector boson production models.
CDF's measured $W+\ge n\;\rm Jets$ cross sections are a factor of 1.7 larger
than the LO QCD prediction for $Q^2_{REN,FAC}=M^2(W)+p_T^2(W)$; the ratio of
the production cross section of $W+\ge1\;\rm Jet$ events to the inclusive $W$
sample ($R_{10}$) agrees with the NLO QCD prediction. 

\section*{Acknowledgments}
%We are grateful to the \D0 and CDF collaborations for discussions
%of their data. 
We acknowledge the support of the US Department of Energy and
the \D0 and CDF collaborating institutions and their funding
agencies in this work.

\end{document}